\documentclass[final,5p,times,twocolumn]{elsarticle}

\biboptions{sort&compress}
\usepackage{amssymb}
\usepackage{amsthm}
\usepackage{amsmath}
\usepackage{multicol}
\usepackage{multirow}
\usepackage{siunitx}
\usepackage{xparse}
\usepackage{booktabs}
\usepackage{graphicx}
\usepackage{svg}
\usepackage{enumitem}
\usepackage{tabularx} % for 'tabularx' env. and 'X' col. type

\journal{Neurocomputing}

\begin{document}
\begin{frontmatter}

%% Title, authors and addresses

%% use the tnoteref command within \title for footnotes;
%% use the tnotetext command for theassociated footnote;
%% use the fnref command within \author or \affiliation for footnotes;
%% use the fntext command for theassociated footnote;
%% use the corref command within \author for corresponding author footnotes;
%% use the cortext command for theassociated footnote;
%% use the ead command for the email address,
%% and the form \ead[url] for the home page:
%% \title{Title\tnoteref{label1}}
%% \tnotetext[label1]{}
%% \author{Name\corref{cor1}\fnref{label2}}
%% \ead{email address}
%% \ead[url]{home page}
%% \fntext[label2]{}
%% \cortext[cor1]{}
%% \affiliation{organization={},
%%            addressline={}, 
%%            city={},
%%            postcode={}, 
%%            state={},
%%            country={}}
%% \fntext[label3]{}

\title{Discriminative Hamiltonian Variational Autoencoder for Accurate Tumor Segmentation in Data-Scarce Regimes}

%% use optional labels to link authors explicitly to addresses:
\author{Aghiles Kebaili$^a$}

\author{Jérôme Lapuyade-Lahorgue $^a$}

\author{Pierre Vera $^{a,b}$}

\author{Su Ruan$^a$}

\affiliation{organization={LITIS UR 4108, University of Rouen-Normandy},
            city={Rouen},
            postcode={76000},
            state={Normandy},
            country={France}}

\affiliation{organization={CLCC Henri Becquerel},
            city={Rouen},
            postcode={76038},
            state={Normandy},
            country={France}}

\begin{abstract}
Deep learning has gained significant attention in medical image segmentation. However, the limited availability of annotated training data presents a challenge to achieving accurate results. In efforts to overcome this challenge, data augmentation techniques have been proposed. However, the majority of these approaches primarily focus on image generation. For segmentation tasks, providing both images and their corresponding target masks is crucial, and the generation of diverse and realistic samples remains a complex task, especially when working with limited training datasets. To this end, we propose a new end-to-end hybrid architecture based on Hamiltonian Variational Autoencoders (HVAE) and a discriminative regularization to improve the quality of generated images. Our method provides an accuracte estimation of the joint distribution of the images and masks, resulting in the generation of realistic medical images with reduced artifacts and off-distribution instances. As generating 3D volumes requires substantial time and memory, our architecture operates on a slice-by-slice basis to segment 3D volumes, capitilizing on the richly augmented dataset. Experiments conducted on two public datasets, BRATS (MRI modality) and HECKTOR (PET modality), demonstrate the efficacy of our proposed method on different medical imaging modalities with limited data.

% In this paper, we present a hybrid architecture that combines the Hamiltonian Variational Autoencoder (HVAE) with a discriminative regularization approach. Our enhanced HVAE provides a more accurate estimation of the posterior distribution, resulting in the generation of realistic medical images with reduced artifacts and out-of-manifold generations. 

% Subsequently, we leverage the joint generation of medical images and segmentation masks from the improved HVAE to train a 2D U-Net architecture. This architecture operates on a slice-by-slice basis to segment 3D volumes, capitalizing on the richly augmented dataset. Through extensive experiments conducted on two publicly available datasets, namely the MICCAI's Brain Tumor Segmentation Challenge (BRATS) and the Head and Neck Tumor Segmentation Challenge (HECKTOR), we demonstrate the efficacy of our proposed method on different medical imaging modalities. Our approach exhibits improvements in image quality, fine-grained tumor mask generation, and addresses challenges posed by limited training data.

\end{abstract}

%%Graphical abstract
% \begin{graphicalabstract}
%\includegraphics{grabs}
% \end{graphicalabstract}

%%Research highlights
% \begin{highlights}
% \item Research highlight 1
% \item Research highlight 2
% \end{highlights}

\begin{keyword}
%% keywords here, in the form: keyword \sep keyword
Deep learning \sep Data augmentation \sep Tumor segmentation \sep Generative modeling \sep Variational Autoencoder \sep MRI \sep PET \sep Hamiltonian Variational Autoencoder

%% PACS codes here, in the form: \PACS code \sep code

%% MSC codes here, in the form: \MSC code \sep code
%% or \MSC[2008] code \sep code (2000 is the default)

\end{keyword}

\end{frontmatter}

%% \linenumbers

% For instance, consider the case of retinal fundus image dataset used for diabetic retinopathy detection. If data augmentation applies excessive rotations or translations, it may introduce unrealistic optic disc positions or retinal vessel distortions, confounding subsequent analysis and decision-making, this illustrates the potential drawbacks of data augmentation.

%% main text
\section{INTRODUCTION}
Deep learning has witnessed remarkable advancements in the field of medical imaging, demonstrating successful outcomes in various segmentation tasks across diverse modalities, including Magnetic Resonance Imaging (MRI) \cite{lundervold2019overview,huang2023semi}, Positron Emission Tomography (PET) \cite{islam2020gan,huang2022lymphoma}, and Computed Tomography (CT) \cite{song2021deep,trullo2019multiorgan}. However, the limited availability of annotated medical imaging data poses a significant challenge, stemming from the rarity of certain pathologies and rigorous medical privacy regulations. Consequently, manual delineation of tumor masks by physicians becomes a laborious and time-consuming process. In this context, data augmentation has emerged as an inseparable part of deep learning, enabling models to overcome the limitations of low sample size regimes, and generalize better \cite{krizhevsky2017imagenet}. Data augmentation encompasses a set of trivial transformations, such as rotations, noise injection, random cropping and more, applied to augment the training dataset. Nevertheless, when dealing with the inherent complex structures in medical images, conventional augmentation operations may introduce image deformations and generate aberrant data, leading to misrepresentations. Thus, advanced deep learning-based data augmentation techniques have been proposed to address these challenges, aiming to generate synthetic samples that closely resemble real data while preserving the semantic integrity of medical images \cite{kebaili2023deep, song2021deep, krizhevsky2017imagenet}. Generative adversarial networks (GANs) \cite{goodfellow2014generative} have found wide application in the field of medical imaging \cite{wang2023fedmed,trullo2019multiorgan} and have been recommended in numerous literature reviews on data augmentation due to their ability to generate realistic images \cite{chen2022generative}. However, GANs exhibit certain limitations, including learning instability, convergence difficulties, and the well-known issue of mode collapse \cite{mescheder2018training}, wherein the generator produces a restricted range of samples. The complex architecture of GANs, which relies on a two-player adversarial game between a generator $G$ and a discriminator $D$, contribute to these challenges. Adversarial training processes do not impose a strong enough constraint for a precise estimation of the target probability density \cite{bond2021deep}, as Nash Equilibrium is hard to achieve, and may demand large amounts of data to better represent the target distribution \cite{li2022comprehensive}. This mismatch between training and target distributions results in a mode collapse. Some studies have explored the use of semi-supervised learning \cite{odena2016semi} to improve GAN convergence, since acquiring unlabeled data is more feasible. This approach faces challenges in the context of segmentation tasks, since it requires generalizing both the image and the target mask, making the application of semi-supervised learning less feasible.

As an alternative, variational autoencoders (VAEs) \cite{kingma2013auto}, though historically less explored, have increasingly gained attention in recent years as a data augmentation approach, outperforming GANs in terms of output diversity and mode coverage \cite{xiao2021tackling}. Moreover, one important advantage of VAEs that is less discussed in the litterature is their capacity to operate better with smaller datasets \cite{chadebec2021data,dai2023swin,kebaili2023end}. This advantage can be attributed to the autoencoding nature of their architecture (refer to Section \ref{sec3}). Furthermore, VAEs are based on a likelihood optimization algorithm, which enables a more precise estimation of the probability density function of the data \cite{bond2021deep}. Despite their potential, VAEs present a significant challenge; their tendency to often produce blurry and hazy output images. This undesirable effect is primarily attributed to the regularization term in the loss function. To address this limitation, recent research has focused on developing various methods to enhance VAEs performance \cite{iafvae,razavi2019generating}. These methods aim to mitigate the blurriness and improving the overall fidelity of the generated images. These methods can be categorized into three distinct approaches. The first approach entails employing more complex prior distributions beyond the standard Gaussian distribution \cite{dilokthanakul2016deep,tomczak2018vae}. The second approach focuses on improving the approximation of the posterior distribution in the Evidence Lower Bound (ELBO), leading to a more accurate representation of the real data distribution \cite{kingma2016improved,rezende2015variational}. The third approach, although less explored, involves modeling the latent space using non-Euclidean geometries that are not vector spaces, such as Riemannian geometries \cite{chadebec2022data,davidson2018hyperspherical}. Additionally, emerging techniques such as diffusion models \cite{ho2020denoising,kazerouni2022diffusion} are gaining recognition for their ability to generate high-quality images. However, their high sampling time, considerable computational demands, and high memory costs make them hardly usable for data augmentation, especially in data-scarce regimes. As an example, the work in \cite{dorjsembe2023conditional} is indeed in 3D but demands substantial memory since generation occurs in pixel space with a Denoising Diffusion Model \cite{ho2020denoising}. Moreover, we've essentially seen in recent years studies focusing on generating 2D images of different medical imaging modalities \cite{stojanovski2023echo,shrivastava2023nasdm,chen2023berdiff}. One might consider using these methods for a 2D-to-3D approach; however, the sampling time of such methods makes the 2D-to-3D approach impractical, as 2D diffusion would take excessively long to generate a volume in a slice-by-slice manner \cite{han2023medgen3d}. The Latent Diffusion Model (LDM) \cite{rombach2022high} has emerged as an alternative to pixel-space diffusion models, functioning on a learned low-dimensional latent representation of the data. While these models partially reduce memory usage and computational expenses linked with diffusion models, they still require a pretrained autoencoder to reconstruct the latent variables into pixel-space. Yet, training such autoencoders in data-scarce scenarios is challenging due to high overfitting risks, especially for high-resolution 3D volumes.

While recent studies have primarily focused on image generation or translation \cite{gan2022esophageal, liang2021data}, in the context of tumor segmentation, generating images alone is insufficient \cite{niyas2022medical}. It becomes crucial to generate both images and their corresponding tumor masks, which serve as ground truth for segmentation tasks. Unlike GANs, few studies have explored the use of VAE-based architectures for data augmentation, primarily due to the fuzzy nature of the generated images. Existing studies often resort to hybrid architectures, combining VAEs with GANs to improve image quality \cite{kebaili2023deep}. Nevertheless, these hybrids introduce complexity and may demand extensive hyperparameter tuning to mitigate adversarial training instabilities, potentially requiring larger datasets \cite{li2022comprehensive}. Alternatively, some works leverage conditional VAEs \cite{pesteie2019adaptive,zhuang2019fmri} for an improved control over the generation process by conditioning the model on additional information. This approach allows the synthesis of synthetic samples that represent specific subgroups within the data \cite{biffi2018learning,volokitin2020modelling}. However, conditioning on latent variables, as seen in CVAEs, may be suboptimal as it involves providing the mask first, which leads sometimes to a heavy two-stage generation process (see Section \ref{sec31}). Among these, Huo et al.'s work \cite{huo2022brain} stands out, where they propose a Progressive Adversarial VAE architecture. Although innovative, their approach is limited to generating tumor regions and their corresponding masks, which are then inserted directly into authentic images to produce synthetic samples. The masks in this context are given a priori and are not generated simultaneously with the tumors, which makes the generated images less realistic.

% leverages Hamiltonian dynamics to achieve a more accurate approximation of the posterior distribution. This 

In this paper, we present a new approach for the simultaneous generation of medical images and corresponding tumor masks using a Hamiltonian VAE architecture \cite{caterini2018hamiltonian}. The HVAE results in a tighter bound on the log-likelihood of the data and improved generation quality while maintaining a good mode coverage. Our approach is designed to streamline the augmentation process. By generating images and tumor masks concurrently, we aim to eliminate the need for a two-stage generation process, providing high-quality samples to bolster data availability for training segmentation models in scenarios with limited data. To enhance the quality of the generated samples, our approach incorporates additional regularization terms into the loss function of the HVAE. We introduce a discriminative regularization term based on adversarial learning, along with a feature-reconstruction loss term, into the global loss function. Our primary goal is to strike a balance between the core objective function, represented by the loss of the HVAE, and the regularization.  We aim to mitigate the impact on learning stability, particularly in a data-scarce setting, by introducing a parameter $\beta$ as a regularization weight. This effectively improve the visual quality and sharpness of the generated images while preserving the mode coverage of the HVAE. The contributions of this paper are as follows :

\begin{itemize}
\item Designing an auto-encoder with a variational Hamiltonian formulation for its capability to offer more accurate likelihood estimations in data-scarce regimes, with weighted discriminative regularization aimed at improving the quality  of the generated medical images in such regimes.
\item Simultaneous synthesis of medical images and corresponding segmentation masks, facilitating the data augmentation of tumor segmentation tasks and eliminating the need of two-stage generation process.
\item We propose a 2D-to-3D augmentation process given the physical limitations of memory and data requirement for the generation of 3D volumes.
\end{itemize}

This article is organized as follows: Section \ref{sec2} provides a theoretical overview of variational autoencoders. Section \ref{sec3} presents our proposed method for generating medical images and tumor masks. Section \ref{sec4} discusses the experimental evaluation on two datasets. Finally, Section \ref{sec5} concludes the paper.

\section{BACKGROUND \label{sec2}}
\subsection{Variational Autoencoders}
VAEs offer a powerful framework for deriving lower-dimensional latent representations from a given set of observable data $x$. VAEs are likelihood-based generative models and aim to maximize the marginal distribution $p_\theta(x)$, representing the probability of observing the data $x$ given the model parameters $\theta$. Mathematically, the marginal distribution $p_\theta(x)$ can be defined as:

\begin{equation}
p_\theta(x) = \int_{z} p_\theta(x|z)p(z)dz,
\end{equation}
where $p_\theta(x|z)$ is the posterior distribution and $p(z)$ the prior distribution over the latent variables $z$. 
However, directly computing the marginal distribution often becomes intractable due to the necessity of integrating over all possible latent variables. To address this challenge, variational inference is employed to approximate the true posterior distribution $p_\theta(z|x)$ with a more tractable one noted $q_\phi(z|x)$. This formulates the learning objective as minimizing the divergence between the posterior distribution $p_\theta(z|x)$ and $q_\phi(z|x)$, where $\theta$ and $\phi$ represent the parameters of the posterior and variational distributions, respectively. The Kullback--Leibler divergence is commonly used as the measure of this divergence.

\begin{equation} 
\min_{\theta,\phi} D_{KL}(q_\phi(z|x) || p_\theta(z|x)) = \min_{\theta,\phi} \mathbb{E}_{z \sim q_\phi}[\log \frac{q_\phi(z|x)}{p_\theta(z|x)}] \label{kld},
\end{equation} 
\noindent Eq \eqref{kld} can be further simplified by applying the Jensen's inequality :
\begin{equation} 
\begin{aligned}
     \log p_\theta(x) &\geq \underbrace{\mathbb{E}_{z\sim q_\phi}[\log p_\theta(x|z)] - \mathbb{E}_{z\sim q_\phi}[\log \frac{q_\phi(z|x)}{p(z)}]}_{-\mathcal{L}_\text{ ELBO}}.
\end{aligned}
\end{equation}

The Evidence Lower BOund (ELBO) is maximized during the learning process, and it encompasses both a reconstruction term measuring the input data's reconstruction accuracy and a regularization term ensuring the adherence of latent variables to a desired distribution, mostly modeled as Gaussian distribution $\mathcal{N}(\mathbf{0}, \mathbf{I})$. The encoder and decoder, as mapping functions, model $q_\phi(z|x)$ and $p_\theta(x|z)$, respectively. Trained VAEs can synthesize new data points by sampling a random latent vector z from the prior distribution and feeding it to the decoder.

\begin{figure*}[t]
\begin{center}
    \includegraphics[width=\textwidth]{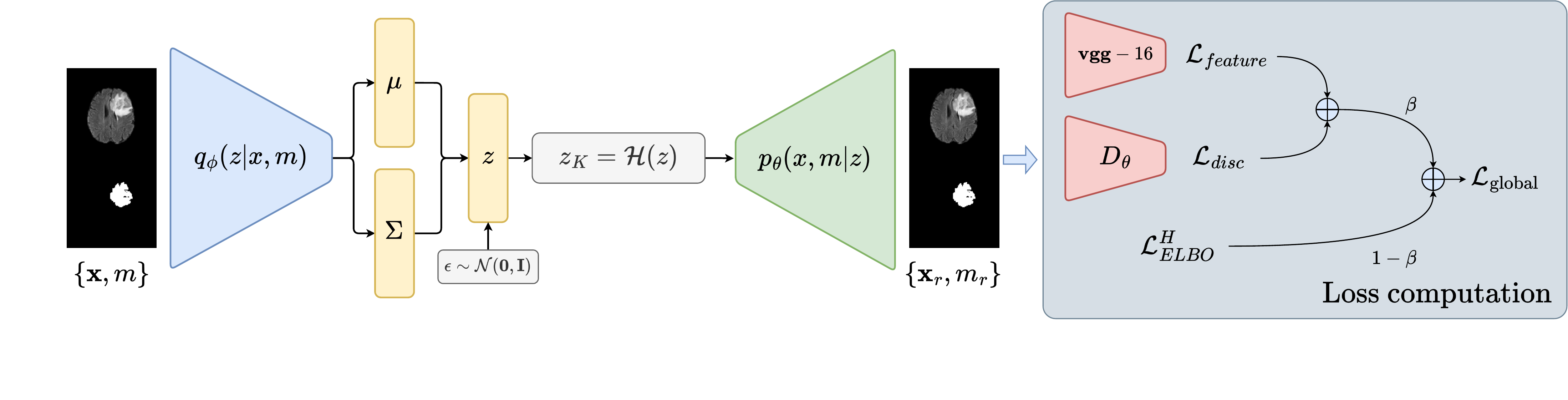}
    \caption{Proposed end-to-end architecture consisting of an encoder $q_\phi$ and decoder $p_\theta$ network, a discriminative regularizer $D_\theta$ , and a pre-trained 16-layer $\mathbf{vgg}$-based architecture. It takes as concatenated multi-channel input the medical image and its corresponding tumor mask noted $\{\mathbf{x}, m\}$ and reconstructs it into $\{\mathbf{x}_r, m_r\}$. Newly generated image pairs are produced by feeding the decoder with a random Gaussian noise vector $z \sim \mathcal{N}(\mathbf{0}, \mathbf{I})$. Further details about the loss computation module can be found in section \ref{sec3}. \label{framework}}
\end{center}
\end{figure*}

\subsection{Hamiltonian Variational Autoencoder}
The Hamiltonian VAE proposed in \cite{caterini2018hamiltonian} is an extension of the standard VAE that utilizes Hamiltonian Monte Carlo (HMC), a Markov chain Monte Carlo method, to improve the generation of images. This method is based on simulating Hamiltonian dynamics on the latent space, which involves treating latent variables as particles moving through a high-dimensional space according to the physical principles of energy conservation and momentum. Hamiltonian dynamics was first introduced in physical mechanics to model total energy as a function of two variables: position and momentum. Hamiltonian differential equations describe how these variables evolve over time. The idea behind HMC is to introduce an auxiliary variable $\rho \sim \mathcal{N}(\mathbf{0}, \mathbf{M})$, known as momentum, to improve the estimation of the posterior distribution. $\mathbf{M}$ is a symmetric and positive definite matrix that can be seen as a mass. By leveraging Hamiltonian dynamics, the HVAE is able to provide a more accurate estimation of the posterior distribution $p_\theta(z|x)$, resulting in higher sampling quality and the avoidance of aberrant generations. The Hamiltonian equation is defined as:
\begin{equation}
\mathcal{H}(z, \rho) = -\log p_\theta(z|x) + \frac{1}{2} \rho^T \mathbf{M}^{-1} \rho \label{eq1},
\end{equation}
where the first term is the potential energy and the second is the kinetic energy. The HMC effectively explores the posterior distribution by iteratively sampling $(z, \rho)$ using Hamiltonian differential equations derived from \eqref{eq1}, leading to the creation of an ergodic and time-reversible Markov chain for $z$ whose stationary distribution corresponds to the target distribution. The diffential equations are :
\begin{equation}
\begin{aligned}
\frac{\partial \rho}{\partial t}(t) = -\nabla_z \mathcal{H}, \;\;\;\;\; \frac{\partial z}{\partial t}(t) = \nabla_\rho \mathcal{H} ,\label{eq2}
\end{aligned}
\end{equation}
where $t$ is the step the the Markov chain and $\nabla_u$ is the respective gradient of $u$. A final proposal $z_K$, where $K$ is the total number of steps, is generated and accepted using the Metropolis-Hastings acceptance ratio \cite{caterini2018hamiltonian}, which ensures that the generated Markov chain has the desired stationary distribution.

\section{PROPOSED METHOD \label{sec3}}
Autoencoding architectures, such as the HVAE, offer a distinct advantage in their ability to operate more effectively with smaller datasets compared to GANs. This advantage can be attributed to the presence of an encoder within the architecture, which serves to extract salient features from input images and represents them in a compact latent space. The encoder’s capacity for identifying these features significantly reduces the search space required for generating new images through the process of reconstruction during training. In contrast, GANs possess a wider search area since it starts directly from the noise, and may encounter challenges in effectively learning features. In essence, autoencoders for generation can be considered as a form of dimensionality reduction, where the representation obtained by the encoder provides an advantageous and strategic starting point for the decoder to accurately approximate the real data distribution. In this context, we propose to use such an architecture to leverage their generative capacity in data-scarce regimes. We present an new end-to-end autoencoding architecture built upon the Hamiltonian module in its latent space, which will offer a distinctive advantage over standard VAEs in approximating data distributions. We further introduce an additional discriminative regularization term into the loss function. Our approach leverages the complementary advantages of HVAEs and adversarial learning to maximize the quality of generated images despite the limited amount of available data. By effectively integrating these two techniques, we strive to achieve the best possible image synthesis performance in the context of data scarcity, while mitigating the risk of mode collapse inherent to adversarial learning through a controlled loss function. To further augment the perceptual realism of generated images, we incorporate a perceptual loss into our global loss function as a third regularization term.  The synthesized images will then be employed for data augmentation for segmentation tasks. Our proposed architecture is illustrated in Figure \ref{framework} and will be expounded upon in detail in the following sections.
% One advantage of the VAE approach is its ability to operate better with smaller datasets compared to GANs. This advantage can be attributed to the presence of an encoder in the VAE architecture, which extracts relevant features from input images for the generation of new ones. The encoder's ability to identify relevant features reduces significantly the search space required for generating new images through the process of reconstruction, unlike GANs, which have a wider search space and may encounter challenges in learning features effectively. In essence, VAEs for generation can be considered as a form of dimensionality reduction. The representation obtained by the encoder provides a better starting point for the decoder to accurately approximate the real data distribution. To this end, we propose an improved architecture for generating medical images that highlights the generation capacity of HVAEs by introducing an additional discriminative regularization term in the loss function. Our approach combines the strengths of VAEs and adversarial learning to improve the quality of generated images while maintaining its ability to learn a compact latent representations and avoid mode collapse. To further improve the perceptual realism of the generated images, we also incorporate a perceptual loss into the global loss function as a third regularization term. In the next section, we will provide a detailed explanation of our proposed architecture which is illustrated in Figure \ref{fig1}.

\subsection{\textbf{Modeling the joint distribution of the images and masks}} \label{sec31}
Conventionally, conditioning labels at the latent space level in generative models offers greater generation freedom. However, this approach presents several drawbacks in the context of segmentation. One of these is the necessity to provide a conditioning mask during the generation, leading to a two-stage generation process where the initial stage entails acquiring the mask. Moreover, learning the conditional distribution can be complex, causing the model to struggle in effectively incorporating condition-specific features into the latent space. As a consequence, suboptimal representations may be generated for corresponding images, resulting in instances where the output does not pixel-perfectly align with the given mask. In the context of image and mask generation, considering the joint distribution $p_\theta(x, m|z)$, where $m$ defines the mask, can lead to more realistic results compared to focusing solely on the conditional distribution $p_\theta(x|z, m)$. By modeling the joint distribution, the generative model can simultaneously consider the dependencies between the images $x$ and the mask $m$, leading to a more comprehensive representation of the data-generation process \cite{stahlschmidt2022multimodal}. This process can be realised by incorporating the mask as an input though a channel-wise concatenation with the corresponding images, as the progressive fusion of the image and mask features from early to deeper layers promotes a more disentangled latent space. This empowers the model to effectively capture the interdependencies and correlations among various data aspects and conditions, potentially leading to more precise and realistic image generations. Additionally, this approach allows us to generate the mask and image pairs in a single pass, without necessitating any preliminary steps or prerequisites. In our specific scenario, through the incorporation of the mask as a secondary input, we can reformulate the Hamiltonian function as follows: 

\begin{equation}
\mathcal{H}(z, \rho) = -\log p_\theta(z|x, m) + \frac{1}{2} \rho^T \mathbf{M}^{-1} \rho \label{eq6}
\end{equation}
An approximation of the differential equations derived from Eq \eqref{eq6} is given through a discretization scheme, namely the leapfrog integrator. The pairs $z, \rho$ are updated using the following scheme :
\begin{equation}
\begin{aligned}
    \tilde{\rho} &= \rho - \frac{\epsilon}{2} \odot \nabla_z U_\theta(z|x, m),\\
    z' &= z + \epsilon \odot \tilde{\rho},\\
    p' &= \tilde{\rho} - \frac{\epsilon}{2} \odot \nabla_z U_\theta(z'|x, m),
\end{aligned}
\end{equation}
where $\epsilon \in \mathbb{(R^+)}^l$ are the individual leapfrog step sizes per dimension, $l$ is the dimension of the momentum, $\odot$ is the elementwise multiplication operation, and $U_\theta(z|x, m)$ corresponds to the potential energy term. 
The HVAE’s loss function, denoted as $\mathcal{L}_{\text{ELBO}}^{\text{H}}$, is constructed upon the HMC module and includes additional analytical evaluations for certain terms \cite{caterini2018hamiltonian}. After considering the input mask, we can define the loss function as follow :
\begin{equation}
\begin{aligned}
\mathcal{L}_{\text{ELBO}}^{\text{H}} &= \mathbb{E}_{z_0 \sim q_{\phi}} [\log p_\theta(x, z_K, m) - \frac{1}{2}\rho_K^T \mathbf{M}^{-1} \rho_K\\ &- \log q_{\phi}(z_K|x, m)] \label{eq3},
\end{aligned}
\end{equation}
where $\log q_{\phi}(z_K|x) \sim \mathcal{N}(z_K; \mathbf{0}, \mathbf{I})$ is the log-likelihood of the latent vector $z_K$. The term $\log p_\theta(x, z_K, m)$ can be further decomposed into two distinct components: $\log p_\theta(x, m | z_K)$ and $\log p_\theta(z_K)$. The first component, $\log p_\theta(x, m | z_K)$, denotes the reconstruction process for both the image and its associated mask, given on the latent vector $z_K$. In order to address the mask reconstruction, we augment the loss function by including an additional term that involves the calculation of cross-entropy between the reconstructed mask and the ground truth mask. On the other hand, the second component, $\log p_\theta(z_K)$, corresponds to the log-probability of the vector $z_K$, representing its likelihood.

\subsection{\textbf{Feature-reconstruction regularization}}
Pixel-reconstruction losses play a crucial role in maintaining global structures in the generated images, but they often suffer from the trade-off between preserving fine details and producing blurry outputs. Moreover, these losses fail to capture the subtleties of human perception, such as texture details, which are important for accurate tumor identification. To address these limitations, we introduce a regularization term inspired by the work of Johnson et al. \cite{johnson2016perceptual}. Instead of enforcing exact pixel-level correspondence between the output image $\hat{y}$ and the target image $y$, our approach focuses on achieving similar feature representations. This technique, initially proposed for style transfer, proves to be effective in enhancing the structure, details, and realism of tumor images, as tumors are the regions of interest in our context. We propose the utilization of this regularization based on the ReLU activation layers of a pre-trained 16-layer VGG network \cite{simonyan2014very}. By utilizing these intermediate feature representations, we encourage the output image to have similar high-level features to the target image. The feature reconstruction loss $\mathcal{L}_{\text{feature}}(\hat{x}, x)$ is defined as :

\begin{equation}
\mathcal{L}_{feature}(\hat{x}, x) = \sum^L_j ||\phi_j(\hat{x}) - \phi_j(x)||^2_2,
\end{equation}
where $\phi_j(x)$ is the feature map of the $j$-th layer of the network $\phi$ for a given input image $x$, and $\phi_j(\hat{x})$ for the reconstructed image. $L$ is the total number of layers.

The feature reconstruction loss is computed progressively for both early and later layers of the network. Minimizing this loss on early layers ensures visually indistinguishable images, while maintaining similar spatial structures to real images on later layers. In addition to the perceptual loss, we incorporate a pixel-wise loss to ensure that the generated images closely match the pixel values of the target images. Although our primary focus is on reconstructing the original image, the pixel-wise loss, implemented as an L1 loss, helps enhance image sharpness and compensates for the loss of fine details caused by the cross-entropy loss of $\mathcal{L}_{\text{ELBO}}^{\text{H}}$.  The total feature-reconstruction loss becomes as :
\begin{equation}
    \mathcal{L}_{\text{rec}}(\hat{x}, x) = \mathcal{L}_{\text{feature}}(\hat{x}, x) + \frac{1}{N} \sum_{i=1}^N ||\hat{x}_i - x_i ||_1
\end{equation}

By combining the feature reconstruction loss and the pixel-wise loss, we achieve a balance between capturing high-level features and preserving pixel-level similarity during reconstruction, therefore improving the quality of the generated images.

\begin{figure*}[]
\begin{center}
    \includegraphics[width=\linewidth]{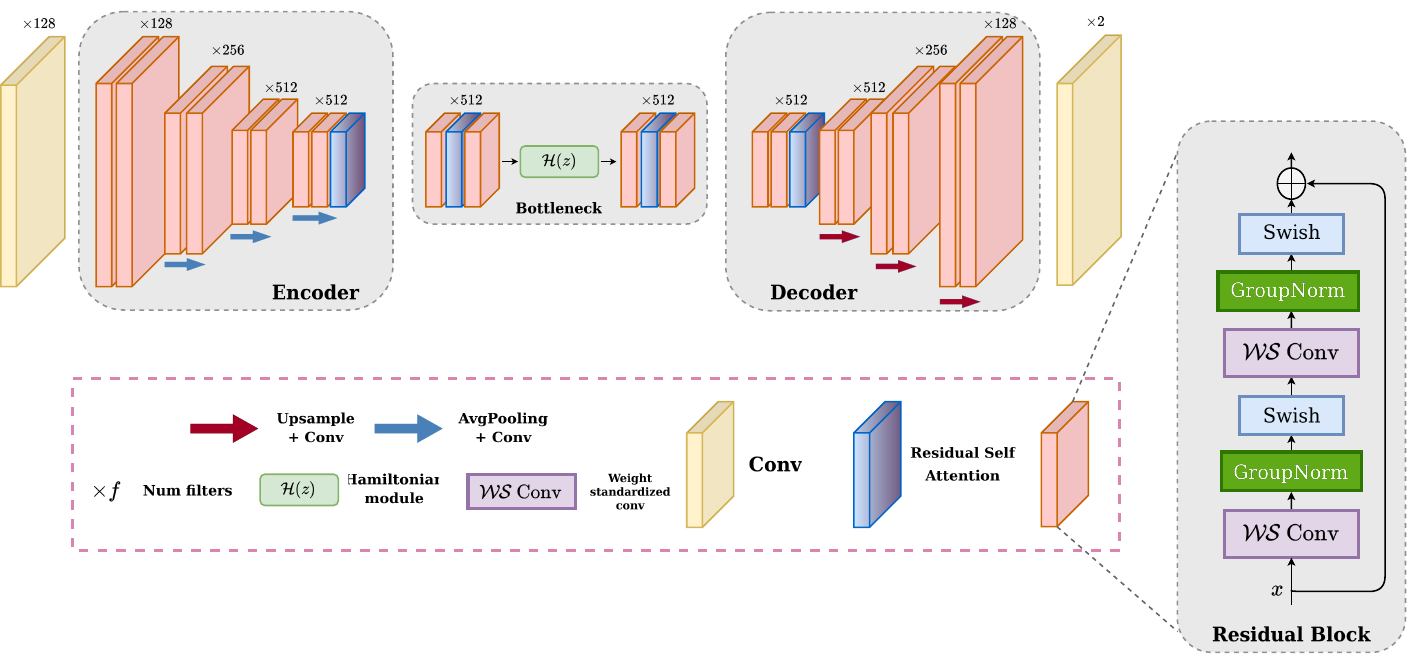}
    \caption{Detailed illustration of our proposed Hamiltonian autoencoding architecture for medical image and mask generation. More details are specified in section 3.6.\label{archi}}
\end{center}
\end{figure*}

\subsection{\textbf{Discriminative regularization}}
Our proposed method harnesses the power of adversarial learning by incorporating a discriminative regularization, partially addressing the issue of low-quality and fuzzy image generation encountered with traditional VAEs. This approach allows us to improve the visual quality and sharpness of the generated images while preserving the diversity of generated samples.

The discriminative regularization term is defined as follows:
\begin{equation}
\begin{aligned}
\mathcal{L}_{\text{disc}} &= \mathbb{E}_{x \sim p_{\text{data}}(x)}[\log D_{\theta}(x)] \\
&+ \mathbb{E}_{z \sim p(z)}[\log(1-D_{\theta}(G_{\phi}(z)))],
\end{aligned}
\end{equation}
where $D_{\theta}$ represents the discriminator network, $G_{\phi}$ represents the generator network, $p_{\text{data}}(x)$ is the data distribution, and $p(z)$ is the prior distribution of the latent space.

To maintain a stable training process, the regularization term is associated with a small coefficient $\beta$, mitigating any potential instabilities that may arise from the adversarial training process. This ensures that the primary objective remains the minimization of the evidence lower bound $\mathcal{L}_\text{ELBO}^\text{H}$ in a traditional training fashion. Furthermore, the discriminative regularization is introduced into the global loss only after a certain number of iterations. This delay allows the network to learn the intrinsic features of the data distribution effectively, avoiding conflicts between the two learning objectives and promoting a smoother convergence. The global loss function can be then summerized to :

\begin{equation}
\begin{aligned}
\mathcal{L}_{\text{global}} = \alpha \cdot \mathcal{L}_{\text{ELBO}}^{\text{H}} + \beta \cdot (\mathcal{L}_{\text{disc}} + \mathcal{L}_{\text{rec}}),
\end{aligned}
\end{equation}
where $\alpha \in [0, 1]$ and $\beta = 1 - \alpha$, allowing for the adjustment of the relative importance between $\mathcal{L}_\text{ELBO}^\text{H}$ and the regularization terms.

\subsection{\textbf{Network architecture}}
As depicted in Figure \ref{archi}, our architecture presents an end-to-end convolutional autoencoding network, combining an encoder and decoder that are mirror images of each other. The architecture is designed with four encoding and decoding blocks, each comprising a weight standardized convolutional layer, GroupNorm, and Swish activation. These blocks incorporate residual connections, allowing information flow from the input to the output of each block, which aids in preserving valuable features during the encoding and decoding processes. The convolutional layers within each block employ a 3$\times$3 kernel and adjust the number of filters based on the block's depth, ultimately reaching 512 filters. Notably, the first and last blocks of the encoder and decoder, respectively, are equipped with residual self-attention layers, enhancing the model's ability to focus on relevant features and capture long-range dependencies. To efficiently handle the latent space, the bottleneck is fully convolutional, and the mean and log variance vectors are modeled as matrices. Additionally, an attention module \cite{vaswani2017attention} is integrated into the latent space, allowing the model to highlight the most critical features for accurate image reconstruction. Notably, this design choice contributes to a lightweight architecture compared to other generative models, totaling merely 23 million parameters.

The network takes a concatenated channel-wise input of the mask and the image in order to model the joint distribution. Although a multi-branch model with multiple encoders could have been beneficial, it was not feasible due to the computational complexity of Hamiltonian Monte Carlo, which requires multiple passes through the decoders, significantly prolonging the learning process. For loss computation, the perceptual loss is calculated using specific layers (2, 7, 12, and 21) of the VGG16 network. Moreover, the discriminator is designed as a Fully PatchGAN \cite{isola2016image}, contributing to the adversarial learning framework's stability.

\section{EXPERIMENTS \label{sec4}}
In this study, we conducted a comprehensive comparative analysis involving our proposed architecture and various competing generative models as data augmentors for the tumor segmentation task. We present both qualitative and quantitative results, examining the quality of synthesized images in the context of two distinct medical imaging modalities. Furthermore, we further evaluate the models' performance on the tumor segmentation task to assess their effectiveness and robustness as data augmentors.

\subsection{\textbf{The 2D/3D Trade-Off}}
Our approach involves segmenting 3D volumes in a slice-by-slice manner, driven by the difficulties in training 3D models in medical imaging. Training 3D generative models requires substantial amounts of data, which is often scarce in the medical field. The three-dimensional nature of the data provides valuable contextual information for solving downstream tasks. However, learning three-dimensional kernels is inherently more complex than in 2D due to the intricate patterns requiring more observable data points for a better recognition. It also involves excessive computational demands, further exacerbating the challenge. Sacrificing the three-dimensional aspect of the data reduces the complexity of the training process and provides more observable data points to train robust 2D generators, when deconstructed into slices. This yields an acceptable compromise for enhancing segmentation tasks in a straightforward manner. This holds particularly true in medical imaging, where most modalities share common features across slices. By focusing on 2D slices and disregarding the volumetric aspect, we gain several advantages. Firstly, we significantly simplify the generation process, allowing the training of a generative model capable of producing images and masks with satisfactory quality under data-scarce conditions. Secondly, we augment the available data samples by generating new ones, crucial for training robust segmentation models. This 2D-centric approach also facilitates the utilization of existing 2D segmentation architectures, which are well-established and readily available.

Our segmentation process is streamlined by initially selecting slices containing tumors, breaking down the 3D problem into a series of manageable 2D tasks. In practical scenarios, a medical operator can interact minimally by identifying the starting and ending levels of the tumor, similar to what has been proposed in \cite{ma2024segment}. Alternatively, an automated approach can be employed using a tumor classifier to identify slices with tumors, offering a fully automated segmentation process. Our experimentation showcases the superiority of our proposed 2D-to-3D approach over standard 3D segmentations in data-scarce scenarios. The simplicity and efficiency of our approach make it particularly suitable for medical imaging scenarios with limited data availability, offering a practical and effective solution for improving segmentation tasks while demanding fewer data and computational resources.

\subsection{\textbf{Datasets}}
We evaluate the efficacy of our proposed method using two widely-used publicly available datasets: the BRAin Tumor Segmentation (BRATS2021) dataset \cite{baid2021rsna} and the HEad and neCK tumOR segmentation (HECKTOR2022) dataset \cite{andrearczyk2021overview}. These datasets cover two distinct medical imaging modalities, namely MRI and PET. By conducting experiments on diverse imaging modalities, we thoroughly examine and validate the effectiveness of our approach across different data domains.

The BRATS2021 dataset contains 1258 subjects comprising high-grade glioma (HGG) low-grade glioma (LGG) cases, each consisting of four MRI modalities: T1, T1c, T2, and FLAIR. The images are provided in the NIfTI format, with a volume shape of 240$\times$240$\times$155 and a voxel resolution of 1$\times$1$\times$1 $mm^3$. The images are skull-stripped and co-registered to the same anatomical template. The ground truth segmentation masks are provided for the tumor core (TC), enhancing tumor (ET), and whole tumor (WT) regions forming three tumor labels. For our experiments, we only consider the WT region.
The HECKTOR2022 dataset, on the other hand, is a publicly available dataset focused on the segmentation of head and neck tumors. It consists of a diverse set of 882 subjects, where each subject's data includes PET and CT images. The PET images provide functional information, while the CT images offer anatomical details, each with a volume shape of 256$\times$256$\times$256 and a voxel resolution of 1$\times$1$\times$1 $mm^3$. For the segmentation task in HECKTOR2022, the ground truth annotations are provided for the primary Gross Tumor Volume (GTVp) and nodal Gross Tumor Volume (GTVn). In our context, we only consider the GTVp region. The PET scans can have lower and variable dimensions. To ensure the consistency of the data, we used tools from TorchIO library to standardize and align the entire dataset to a fixed size of 256$\times$256$\times$256 around tumors.
We exclusively employ the FLAIR modality from the BRATS2021 dataset and the PET modality from the HECKTOR2022 dataset to evaluate the efficacy of our proposed method. We chose FLAIR due to its adequacy for training tumor segmentation tasks, while for HECKTOR, we only use PET modality for several reasons: Firstly, the absence of injected contrast agents during the acquisition of CT images makes tumors challenging to distinguish from soft tissues, as both exhibit similar densities. In the medical field, only contrastive CT scans are used for delineating tumors. Moreover, the segmentation masks provided are derived solely from PET scans. Attempting to use these masks exclusively with CT scans is inaccurate due to differences in voxel spacing. The diverse tumor representations offered by PET and MRI provide a robust foundation to thoroughly evaluate the effectiveness of our proposed methods.

\subsection{\textbf{Training settings}}

In order to simulate a data-scarce scenario, we intentionally limit our initial training set to 30 subjects from a total of 1258 subjects for BRATS dataset, and 882 subjects for HECKTOR. Upon deconstructing this set into 2D slices, we obtain a final training set of 771 images for BRATS and 267 for HECKTOR, accompanied by their corresponding masks, which will be employed for training our generative models. The results of the generation task are elaborated in Section 4.6.2. For our segmentation task, we adopt a 5-fold cross-validation strategy, constructing each training fold with 30 randomly selected subjects. Distinct training configurations are created by augmenting these training sets with varying quantities of synthetic images, with each synthetic set being unique from one configuration to another. The evaluation encompass subjects that were not included in the training set, comprising a total of 1228 subjects for BRATS and 852 subjects for HECKTOR. We record the mean and standard deviation obtained for each experiment, ensuring a robust evaluation of results. This approach guarantees result reliability, enabling us to systematically assess the impact and contribution of synthetic images while maintaining methodological consistency. Its implementation mitigates potential biases introduced by real images, allowing for a focused analysis of the synthetic images and their overall robustness. Further elaboration on the augmentation process can be found in Section 4.7.
%----

We devote a part of section 4.6 to studying the impact of the number of training volumes used on the final results. This investigation is crucial for understanding the model's behavior when exposed to varying amounts of training data.

% To confirm this hypothesis, we trained a 3D U-Net on the limited dataset and quantified the Dice score obtained, resulting in 0.643\% $\pm$ 0.02

% \begin{table*}[]
% \centering
% \begin{tabular}{@{}llccccl@{}}
% \toprule
% \multicolumn{1}{c}{\multirow{2}{*}{$\beta$}} & \multicolumn{3}{c}{BRATS}            & \multicolumn{3}{c}{HECKTOR}          \\ \cmidrule(l){2-7} 
% \multicolumn{1}{c}{}                          & \multicolumn{1}{c}{PSNR} & FID & LPIPS & IS & FID & \multicolumn{1}{c}{LPIPS} \\ \midrule
% $\beta = 0.1$   & 21.427$\pm$2.32 & 92.488$\pm$3.78 & 0.229$\pm$0.01 & 17.422$\pm$1.77 & 102.755$\pm$3.00 & \textbf{0.301$\pm$0.02} \\
% $\beta = 0.01$  & \textbf{24.525$\pm$1.55} & \textbf{88.299$\pm$3.03} & \textbf{0.227$\pm$0.02} & \textbf{19.259$\pm$2.12} & \textbf{54.008$\pm$2.47} & 0.309$\pm$0.04  \\
% $\beta = 0.001$ & 23.470$\pm$3.45 & 126.503$\pm$5.69 & 0.235$\pm$0.04 & 17.999$\pm$2.78& 110.349$\pm$1.89 & 0.333$\pm$0.02 \\ \bottomrule
% \end{tabular}%
% \end{table*}

\subsection{\textbf{Evaluation metrics}}
To quantify the capabilities of our proposed method and competing approaches, we divide the evaluation metrics into three distinct groups, each targeting specific aspects of the image synthesis.

\subsubsection{\textbf{Perceptual Metrics for Image Generation:}}
The first category of evaluation metrics aims to assess the quality of the generated images. Perceptual metrics focus on comparing the visual fidelity between real images and their synthetic counterparts.\\

\noindent a. \textit{Peak Signal-to-Noise Ratio (PSNR)}

The PSNR quantifies the reconstruction quality of the generated images concerning the ground truth images. It measures the ratio of the maximum possible power of the images to the power of the noise. Higher PSNR values indicate better image fidelity.
\begin{equation}
    \text{PSNR}(I_{\text{r}}, I_{\text{f}}) = 20 \cdot \log_{10} \left( \frac{{\text{MAX}_I}}{{\sqrt{{\text{MSE}(I_{\text{r}}, I_{\text{f}})}}}} \right),
\end{equation}
where $I_{\text{r}}$ and $I_{\text{f}}$ represent real and generated images, respectively, $\text{MSE}(I_{\text{r}}, I_{\text{f}})$ is the Mean Squared Error between the two images, and $\text{MAX}_i$ is the maximum possible value of $I$.\\

\noindent b. \textit{Frechet Inception Distance (FID)}

The FID \cite{heusel2017gans} measures the similarity between the data distributions of real and synthetic images in the feature space of the pre-trained Inception network \cite{szegedy2015going}.
\begin{equation}
    \text{FID}(P_{\text{r}}, P_{\text{f}}) = \lVert \mu_{\text{r}} - \mu_{\text{f}} \rVert_2^2 + \text{Tr}(\Sigma_{\text{r}} + \Sigma_{\text{f}} - 2(\Sigma_{\text{r}}\Sigma_{\text{f}})^{1/2}),
\end{equation}
where $P_{\text{r}}$ and $P_{\text{f}}$ represent the real and generated image feature distributions, respectively. $\mu_{\text{r}}$ and $\mu_{\text{f}}$ are the mean vectors, and $\Sigma_{\text{r}}$ and $\Sigma_{\text{f}}$ are the covariance matrices of the real and generated features, respectively.\\

\noindent c. \textit{Learned Perceptual Image Patch Similarity (LPIPS)}

LPIPS \cite{zhang2018unreasonable} quantifies the perceptual similarity between the generated and real images based on the distance of deep feature representations extracted from a pre-trained model (AlexNet \cite{alexnet} in our case). Lower LPIPS scores indicate higher image quality and perceptual similarity.

\begin{equation}
\text{LPIPS}(I_r, I_f) = \sum^L_{k=0} \left( \frac{F^k_r}{||F^k_r||_2} - \frac{F^k_f}{||F^k_f||_2} \right)^2,
\end{equation}
where $I_r$ and $I_f$ denote real and generated images, respectively. The terms $F_r^k$ and $F_f^k$ represent their corresponding feature maps obtained from a pre-trained deep neural network across all its layers. The index $k$ signifies the specific layer, and $L$ denotes the maximum number of layers considered.

\subsubsection{\textbf{Distribution Metrics for Mask Generation:}}
The second category of evaluation metrics focuses on assessing the quality of the generated masks based on their data distribution. This entails utilizing divergence metrics such as the Kullback-Leibler Divergence (KLD) and Jensen-Shannon Divergence (JSD), which quantify the disparity between mask probability distributions. For the segmentation task, we use the the Sorensen-Dice Coefficient, also known as the Dice Similarity Coefficient (DSC), which quantifies the overlap between the predicted and ground truth segmentation masks.

\subsection{\textbf{Implementation details}}
The experiments are performed using Python 3.8 and PyTorch 2.0, executed on an Ubuntu 18.04 system with an NVIDIA GeForce RTX A6000 GPU equipped with 48GB of VRAM. We used the AdamW optimizer with moment parameters $\beta_1 = 0.5$ and $\beta_2 = 0.999$ . The learning rate is set to 5e-5, and the weight decay is configured as 1e-6 to mitigate the risk of overfitting, particularly for datasets with limited samples.

\subsection{\textbf{Evaluating quality of synthesized images and masks}}
\subsubsection{Impact of the hyperparameter $\beta$ on generated images}
The hyperparameter $\beta$ directly influences the quality of the generated images. A larger value of $\beta$ introduces more emphasis on the adversarial learning in the overall loss function. As a result, it generates sharper images. However, it is important to note that adversarial learning may require a substantial amount of data before yielding satisfactory results. Therefore, careful consideration of the $\beta$ value is crucial to strike a balance between image quality, mode coverage, and avoiding visual artifacts or degradations that poorly trained adversarial networks can introduce.

\begin{figure}[t]
\begin{center}
\includegraphics[width=\columnwidth]{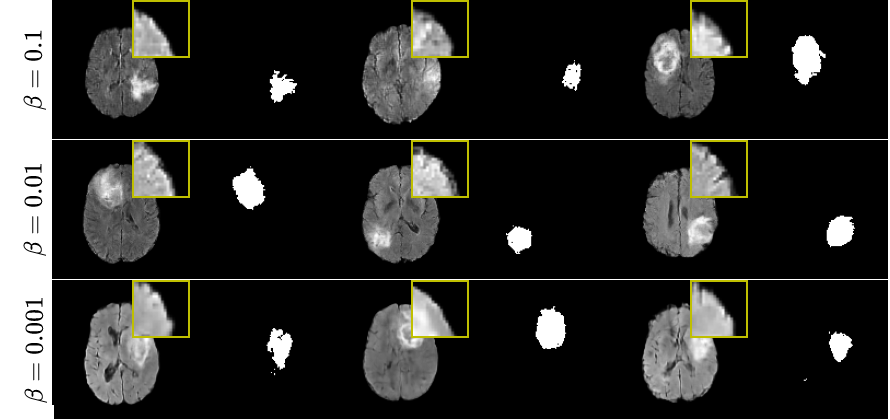}
\caption{Comparison of generated images and masks using our method on the BRATS dataset with varying hyperparameter $\beta$. To gain deeper insights into the image quality and blurriness, we provide a magnified view of the borders, presenting additional details about these specific regions. Each column represent an example of a generated pair of MRI and its corresponding mask.} \label{fig:beta}
\end{center}
\end{figure}

In our experimentation, we explore three orders of magnitude for the $\beta$ value, ranging from 0.001 to 0.1. Beyond a $\beta$ value of 0.1, we observe a negative impact of adversarial learning, leading to overly jagged and less diverse images. On the other hand, a $\beta$ value of 0.001 results in images that are generally blurrier, especially at the edges. Similarly, for $\beta$ = 0.1, where the discriminator's weight is more pronounced, the generated images appear sharper but may exhibit some "checkboard artifacts". To strike a balance between image quality and coverage, we find that a $\beta$ value of 0.01 offers a favorable compromise. In the Table \ref{tab:beta}, we provide a comprehensive summary of PSNR, FID, and LPIPS measures for each $\beta$ value across the two datasets, BRATS and HECKTOR. This table offers insights into the performance of our proposed method and competing approaches with respect to different $\beta$ values.

As it can be seen on the Table \ref{tab:beta}, the impact of $\beta$ is evident in the FID, a metric sensitive to blurry images and various types of artifacts. Consequently, we observe higher FID values for $\beta$ = 0.001 and $\beta$ = 0.1, reflecting the presence of these issues. Figure \ref{fig:beta} presents examples of synthesized images on BRATS using for different $\beta$ values. Notably, for $\beta = 0.001$, the images appear blurrier overall, particularly at the borders. Conversely, with $\beta = 0.1$, the images are sharper but may exhibit some degradations.

\begin{table}[h]
\centering
\resizebox{\columnwidth}{!}{%
\begin{tabular}{@{}llccccl@{}}
\toprule
\multicolumn{1}{c}{\multirow{2}{*}{$\beta$}} & \multicolumn{3}{c}{BRATS}            & \multicolumn{3}{c}{HECKTOR}          \\ \cmidrule(l){2-7} 
\multicolumn{1}{c}{}                          & \multicolumn{1}{c}{PSNR} & FID & LPIPS & PSNR & FID & \multicolumn{1}{c}{LPIPS} \\ \cmidrule(r){1-1} \cmidrule(lr){2-4} \cmidrule(l){5-7}
$\beta = 0.1$   & 15.208 & 92.488 & 0.229 & 18.990 & 102.755 & \textbf{0.301} \\
$\beta = 0.01$  & \textbf{15.622} & \textbf{88.299} & \textbf{0.227} & \textbf{19.570} & \textbf{54.008} & 0.309  \\
$\beta = 0.001$ & 15.409 & 126.503 & 0.235 & 19.203 & 110.349 & 0.333 \\ \bottomrule
\end{tabular}%
}
\caption{Quantitative performance of the proposed method with different $\beta$ values in terms of PSNR (dB $\uparrow$), FID ($\downarrow$) and LPIPS (\%$\downarrow$) on the BRATS and HECKTOR datasets. \label{tab:beta}} 
\end{table}

\subsubsection{Evaluation of the quality of the images}
In this section, we present a comprehensive evaluation of our proposed method in comparison with other competing generative models. The assessment involves both qualitative and quantitative analysis to evaluate the quality of the generated images. The competing models encompass the vanilla VAE architecture \cite{kingma2013auto}, the standard HVAE \cite{caterini2018hamiltonian}, the Least Squares GAN \cite{mao2017least} and the Latent Diffusion Model \cite{rombach2022high}. Additionally, we conducted an ablation study to investigate the impact of augmenting the standard HVAE with a perceptual loss and HVAE coupled with a discriminator. The results of this study help understand the contribution of these modifications to the overall performance. The evaluation outcomes highlight the efficacy of our proposed generator in producing good quality synthetic images with limited training data. Given the table \ref{tab:images}, we observe a gradual improvement going from the standard VAE to our method denoted dHVAE (d stands for \textit{discriminative}). On the other hand, the GAN-based architecture, which is known for yielding superior results to VAE-based architectures \cite{kebaili2023deep} significantly suffers from the scarcity of the data. The LDM benefits from a reconstruction model, affording it greater room to prioritize on image quality. However, a notable drawback encountered during training is the LDM's overfitting, resulting in the reproduction of certain input data identically. Regarding the ablation study, we observe that adding a discriminator or perceptual loss individually to the HVAE does not notably enhance the results compared to the standalone HVAE. However, coupling both the discriminator and perceptual loss through our method, and controlling their contributions with the hyperparameter $\beta$, leads to better performance. During the ablation study, the discriminator and perceptual loss were used with equal weighting to the HVAE, possibly explaining the lack of significant improvement.

\begin{table}[t]
\centering
\resizebox{\columnwidth}{!}{%
\begin{tabular}{@{}llccccl@{}}
\toprule
\multicolumn{1}{c}{\multirow{2}{*}{Methods}} & \multicolumn{3}{c}{BRATS}            & \multicolumn{3}{c}{HECKTOR}          \\ \cmidrule(l){2-7}
\multicolumn{1}{c}{}                          & \multicolumn{1}{c}{PSNR} & FID & LPIPS & PSNR & FID & \multicolumn{1}{c}{LPIPS} \\ \cmidrule(r){1-1} \cmidrule(lr){2-4} \cmidrule(l){5-7}
LSGAN                   & 14.579  & 158.289 & 0.473 & 18.440 & 189.753 & 0.459 \\
LDM  & 15.553  & 91.912 & 0.231 & 19.301 & 83.758  & \textbf{0.289} \\
VAE                     & 15.181  & 106.276 & 0.273 & 19.038 & 90.208  & 0.338 \\
HVAE                    & 15.336  & 124.750 & 0.248 & 19.115 & 74.428  & 0.310 \\
HVAE + disc             & 15.102  & 147.954 & 0.237 & 19.005 & 134.426 & 0.360              \\
HVAE + perc             & 14.923  & 104.767 & \textbf{0.225} & 18.783 & 69.111 & 0.313 \\
dHVAE ($\beta = 0.01$)  & \textbf{15.622} & \textbf{88.299}  & 0.227 & \textbf{19.570} & \textbf{54.008}  & 0.309 \\
\bottomrule
\end{tabular}%
}
\caption{Quantitative performance of the generative models evaluated PSNR (dB $\uparrow$), FID ($\downarrow$) and LPIPS (\%$\downarrow$) on the BRATS and HECKTOR datasets. \label{tab:images}}
\end{table}

\begin{figure*}[h]
\begin{center}
\includegraphics[width=14cm]{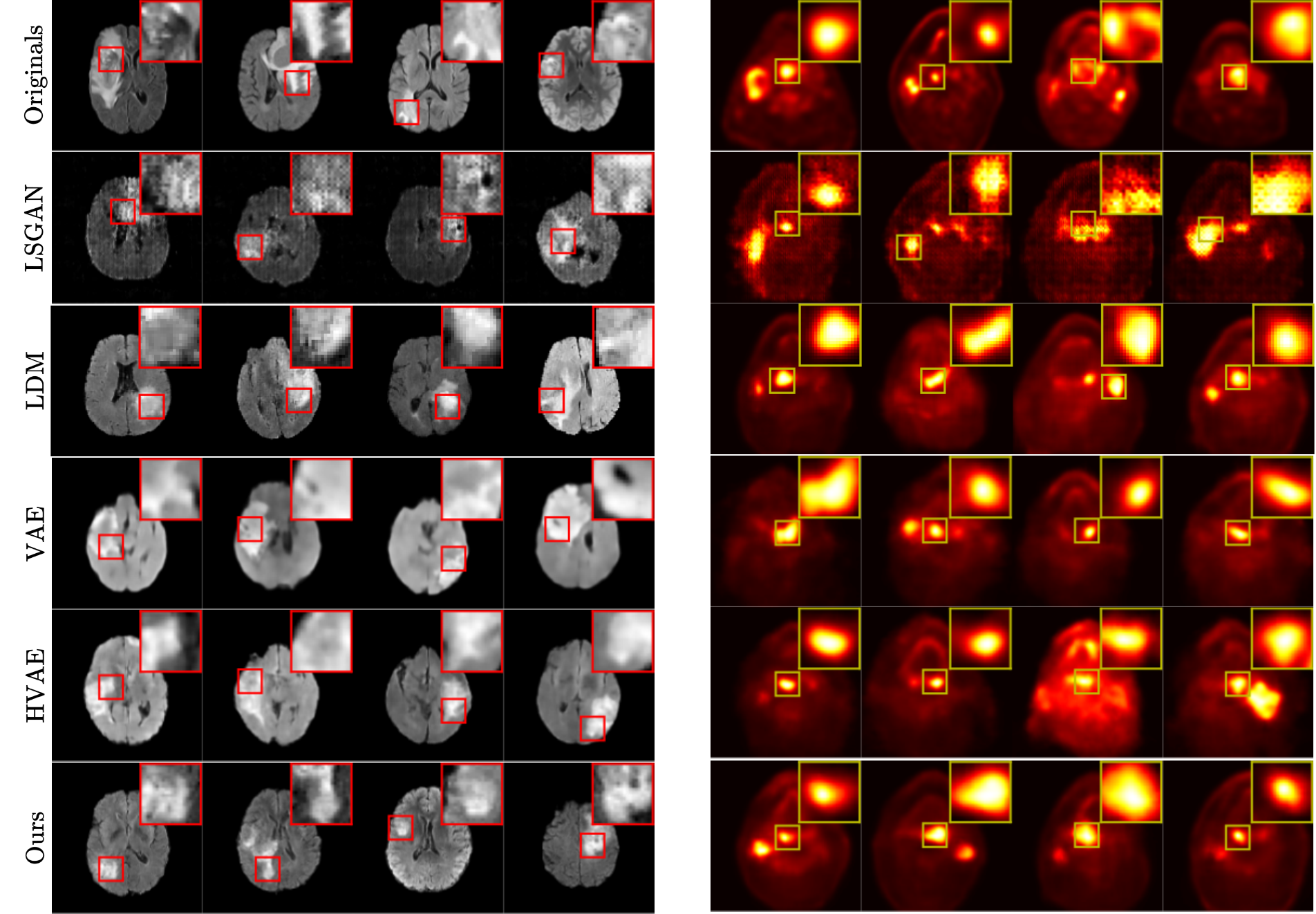}
\caption{Comparison of images generated our proposed method and other generative models on the BRATS and HECKTOR datasets. Each column represents an example of synthetic image produced for both datasets. To provide a closer examination of the tumoral regions texture, a zoomed-in rectangle is presented, offering additional insights into the fine details of these regions.} \label{fig:images}
\end{center}
\end{figure*}

As depicted in Figure \ref{fig:images}, the transition from VAE to our method demonstrates an enhancement in image quality, particularly in representing tumor regions. The LSGAN, on the other hand, exhibits degraded images with visible artifacts. These observations confirm our hypothesis about the potential of autoencoding architecture to effectively handle datasets with limited samples. Figure \ref{fig:images} further illustrates that our method produces the clearest and most realistic images with accurate textures of tumoral regions on the BRATS dataset. Additionally, for the HECKTOR dataset, our method showcases the most realistic images with coherent anatomical structures, capturing the intricate details of PET imaging, while other methods exhibit slightly blurrier images for the VAE-based models and highly degraded images for the GAN-based architecture.

\subsubsection{Evaluation of the quality of the masks}
To assess the quality of the generated masks, we compare the distribution of real masks with that of the generated ones by measuring the divergence between the two distributions. The distributions are quantified by computing the class probabilities for each pixel across a large set of 30 000 synthetic and real masks.
The evaluation metrics employed to analyze the results reveal relatively low scores, owing to the binary and sparse nature of the masks.
We observe from Table \ref{tab:masks} a gradual improvement as we transition from the VAE to our proposed method. Our method demonstrates the best results with the least divergence to the real data distribution compared to other methods, indicating the fidelity of the generated images.

It is important to highlight the significant leap in performance from the VAE-based architectures to the LSGAN. VAE-based architectures inherently optimize the log-likelihood of data, making them well-suited to accurately simulate data distributions. Similarly, the LDM method operates by minimizing the ELBO term, which can be viewed as a likelihood optimization approach. This shared characteristic explains its competitive performance relative to our method. Additionally, since the masks are discrete sparse matrices, there is no concern for image quality and blurriness, making this type of architecture more powerful for mask generation tasks.

\begin{table}[h]
\centering
% \resizebox{8cm}{!}{%
\small
\begin{tabular}{@{}lcccc@{}}
\toprule
\multirow{2}{*}{Methods} & \multicolumn{2}{c}{BRATS}         & \multicolumn{2}{c}{HECKTOR}       \\ \cmidrule(l){2-5} 
                         & JSD             & KLD             & JSD             & KLD             \\ \cmidrule(r){1-1} \cmidrule(lr){2-3} \cmidrule(l){4-5}
LSGAN                    & 0.0049          & 0.0260          & 0.0026          & 0.0104          \\
LDM & 0.0038           & 0.0301         & 0.0016            & \textbf{0.0026} \\
VAE                      & 0.0039          & 0.0217          & 0.0017          & 0.0076          \\
HVAE & 0.0044          & 0.0243          & 0.0025          & 0.0082          \\
dHVAE ($\beta = 0.01$)   & \textbf{0.0032} & \textbf{0.0169} & \textbf{0.0014} & 0.0069          \\ \bottomrule
\end{tabular}
% }
\caption{Quantitative performance of the generated masks evaluated in terms of Jensen-shannon divergence ($\downarrow$) and Kullback-Leibler divergence ($\downarrow$) between simulated distributions of the synthetic masks and the real distribution. \label{tab:masks}}
\end{table}

\subsection{\textbf{Improving segmentation tasks using synthetic images and masks}}
In this section, we address the challenge of improving segmentation tasks through the use of synthetic images and masks as data augmentors. The robust evaluation of data augmentation methods can be complex, as it heavily relies on the number of real and synthetic samples added, which significantly impacts the overall segmentation performance.

\begin{figure}[h]
\begin{center}
\includegraphics[width=\columnwidth]{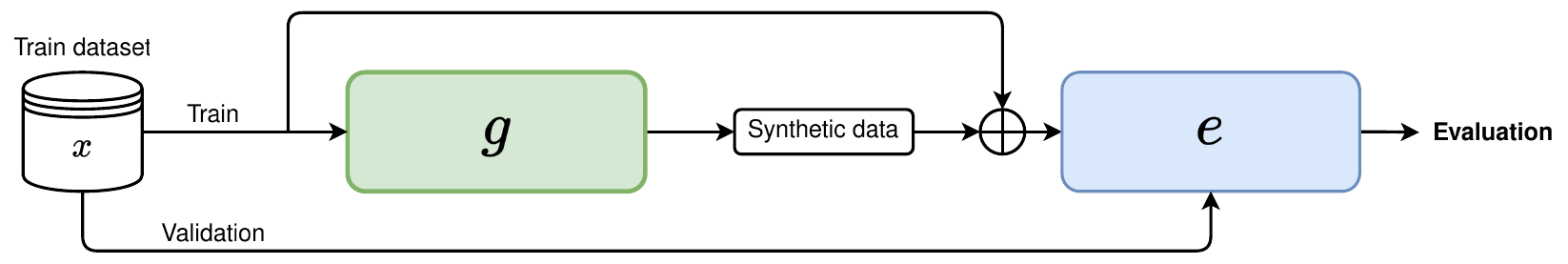}
\caption{Illustration of the augmentation pipeline for a generative-model-based data augmentation. Adapted from \cite{kebaili2023deep}} \label{pipeline}
\end{center}
\end{figure}

The Figure \ref{pipeline} illustrates the pipeline of our generative data augmentation. It involves a generative model $g$ and an evaluation model $e$, which in our case, is a standard U-Net model \cite{ronneberger2015u}. The generator $g$ is trained using a limited training set and then synthesizes additional data samples to augment this same training set. The downstream architecture $e$ is then trained on the augmented training set and evaluated on a separate test set. The reference DSC is obtained from training a conventional U-Net on our limited training set, serving as a benchmark for comparison to assess the impact of synthetic images. Additionally, we present the results obtained from a 3D segmentation to illustrate the challenges encountered in data-scarce scenarios and underscore the advantages of our 2D-to-3D approach in such settings compared to both references. Traditional data augmentation techniques, such as rotation, horizontal/vertical flipping, cropping, etc., are also evaluated and compared to our method with the augmentation factors of $\times$2 and $\times$5. The evaluation of the synthetic images involves varying the number of samples added to the baseline training set, as previously described. By systematically assessing the impact of incorporating synthetic images and masks on the segmentation performance, we aim to identify the optimal augmentation strategy that yields the most significant improvements in accuracy and generalization. The results of our experiments are presented in Table \ref{tab:dice}, providing valuable insights into the improvement achieved through the synthetic data.

\begin{figure*}[t]
\begin{center}
\includegraphics[width=18cm]{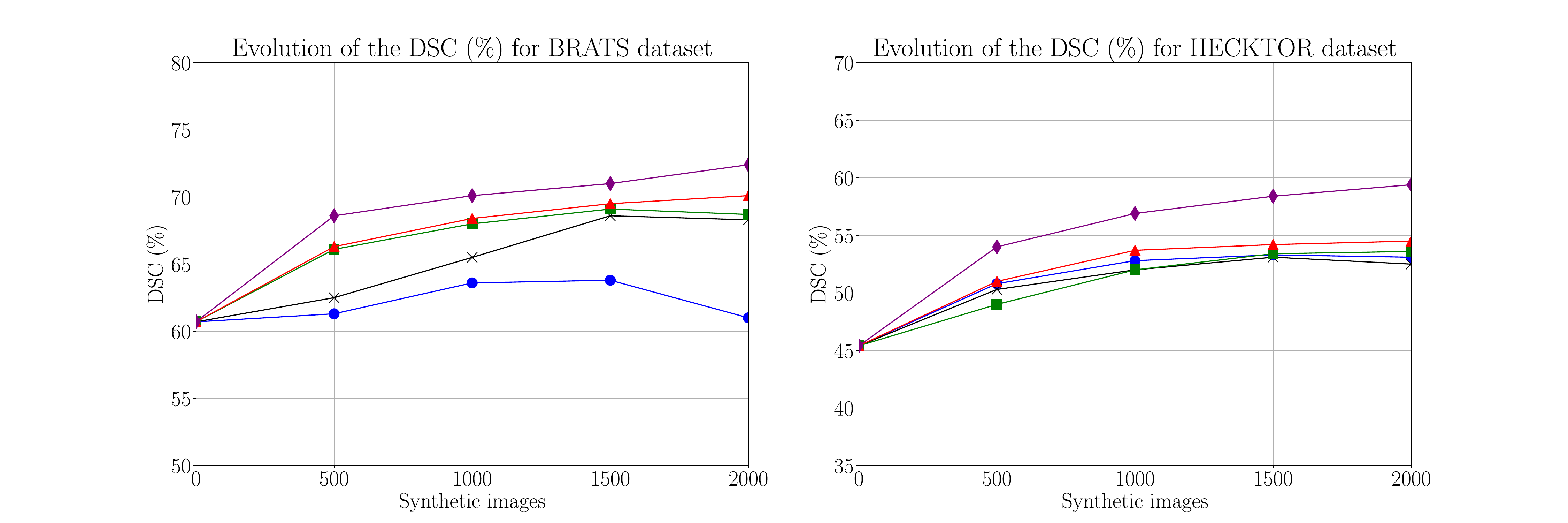}
\caption{Evolution of DSC with the number of synthetic images added. The graph shows the DSC score trends for different configurations, highlighting the impact of synthetic data augmentation on the segmentation performance.} \label{fig:plot}
\end{center}
\end{figure*}

\begin{table}[h]
\centering
\resizebox{7.2cm}{!}{
\begin{tabular}{cccc}
\toprule
\multicolumn{2}{c}{Methods}                                    & BRATS & HECKTOR \\ \cmidrule(r){1-2} \cmidrule(lr){3-3} \cmidrule(l){4-4}
\multicolumn{2}{c}{2D Reference}                           & 0.581$\pm$0.04 & 0.428$\pm$0.03       \\
\multicolumn{2}{c}{Augmented ($\times$2)}                  & 0.630$\pm$0.01 & 0.461$\pm$0.02        \\
\multicolumn{2}{c}{Augmented ($\times$5)}                  & 0.695$\pm$0.01 & 0.563$\pm$0.01        \\
\cmidrule(r){1-2} \cmidrule(lr){3-3} \cmidrule(l){4-4}
\multicolumn{2}{c}{3D Reference}                           & 0.607$\pm$0.01 & 0.454$\pm$0.03       \\ 
\multicolumn{2}{c}{Augmented ($\times$2)}                  & 0.664$\pm$0.01 & 0.533$\pm$0.01        \\
\multicolumn{2}{c}{Augmented ($\times$5)}                  & 0.702$\pm$0.00 & 0.571$\pm$0.01        \\
\cmidrule(r){1-2} \cmidrule(lr){3-3} \cmidrule(l){4-4}
\multirow{4}{*}{LSGAN}                  & + 500 synthetic  & 0.613$\pm$0.04 & 0.508$\pm$0.04 \\
                                        & + 1000 synthetic & 0.636$\pm$0.04 & 0.528$\pm$0.03 \\
                                        & + 1500 synthetic & 0.638$\pm$0.03 & 0.533$\pm$0.03 \\
                                        & + 2000 synthetic & 0.610$\pm$0.01 & 0.521$\pm$0.01  \\
                                        \cmidrule(r){1-2} \cmidrule(lr){3-3} \cmidrule(l){4-4}
\multirow{4}{*}{LDM}                 & + 500 synthetic  & 0.625$\pm$0.01 & 0.503$\pm$0.03 \\
                                        & + 1000 synthetic & 0.655$\pm$0.01 & 0.520$\pm$0.01 \\
                                        & + 1500 synthetic & 0.686$\pm$0.00 & 0.531$\pm$0.01 \\
                                        & + 2000 synthetic & 0.683$\pm$0.00 & 0.525$\pm$0.00\\
                                        \cmidrule(r){1-2} \cmidrule(lr){3-3} \cmidrule(l){4-4}
\multirow{4}{*}{VAE}                    & + 500 synthetic  & 0.661$\pm$0.00 & 0.490$\pm$0.01 \\
                                        & + 1000 synthetic & 0.680$\pm$0.00 & 0.510$\pm$0.01 \\
                                        & + 1500 synthetic & 0.691$\pm$0.01 & 0.534$\pm$0.02 \\
                                        & + 2000 synthetic & 0.687$\pm$0.01 & 0.526$\pm$0.01 \\ 
                                        \cmidrule(r){1-2} \cmidrule(lr){3-3} \cmidrule(l){4-4}
\multirow{4}{*}{HVAE}                   & + 500 synthetic  & 0.663$\pm$0.00 & 0.510$\pm$0.01 \\
                                        & + 1000 synthetic & 0.684$\pm$0.01 & 0.537$\pm$0.01 \\
                                        & + 1500 synthetic & 0.695$\pm$0.00 & 0.542$\pm$0.00 \\
                                        & + 2000 synthetic & 0.701$\pm$0.00 & 0.545$\pm$0.00 \\ 
                                        \cmidrule(r){1-2} \cmidrule(lr){3-3} \cmidrule(l){4-4}
\multirow{4}{*}{Ours ($\alpha = 0.01$)} & + 500 synthetic  & 0.686$\pm$0.01 & 0.540$\pm$0.01 \\
                                        & + 1000 synthetic & 0.701$\pm$0.01 & 0.569$\pm$0.02 \\
                                        & + 1500 synthetic & 0.710$\pm$0.00 & 0.581$\pm$0.00 \\
                                        & + 2000 synthetic & \textbf{0.724$\pm$0.01} & \textbf{0.594$\pm$0.00} \\ 
                                        \bottomrule
\end{tabular}%
}
\caption{Quantitative performance of the generative models is evaluated in terms of DSC (\%↑) on BRATS and HECKTOR. \label{tab:dice}}
\end{table}

Our synthetic samples, generated by our architecture, significantly improve the DSC on both datasets, showcasing notable enhancements of +11.7\% and +14.0\%, respectively, in comparison to the 3D reference. Moreover, the proposed method demonstrates low standard deviation (approximately 1\% DSC), affirming its robustness against data variability. Our proposed architecture also stands out as the only method surpassing traditional augmentation techniques by a factor of 5, underscoring its superiority in enhancing dataset size and diversity. We notice from Table \ref{tab:dice} decline in the improvement of DSC beyond the inclusion of 1500 synthetic images for most methods. This decrease can be attributed to the imbalance between the number of real and synthetic images in the training set. The generated images, while providing augmentation benefits, have not yet reached a level of indistinguishability from real images. The unbalanced ratio between real and synthetic images introduces a bias that causes the model to primarily optimize the loss through synthetic images, negatively impacting the DSC score on the test set and limiting the contribution of more synthetical images. However, it is important to highlight that the DSC does not exhibit a decrease in the case of our proposed method. In the context of the HECKTOR dataset, the earlier decline in DSC scores for VAE and LSGAN methods can be attributed to the inherent challenges associated with segmenting PET imaging data. Specifically, PET images may present instances of metabolic activity in healthy organs, which can be mistaken for tumors during the segmentation, thus further complicating the segmentation task. For the LDM, we observe a stagnation in the segmentation results (Table \ref{tab:dice}), despite its superior image quality compared to VAE and HVAE (Table \ref{tab:images}). This is potentially due to its overfitting, resulting in redundancies in the training set and less diversity in the generated images. Figure \ref{fig:plot} depicts the evolution of the DSC as a function of the number of synthetic images added, presented in the form of a curve graph. To gain a deeper understanding of the impact of synthetic images on U-Net's learning, we compile the results of several segmentations in Table \ref{tab:volumes}, varying this time the number of the real volumes used for each training. Additionally, we present the results of segmentation by combining our method with traditional data augmentation using a factor of $\times5$. We incorporate 2000 synthetic images for both BRATS and HECKTOR datasets. These numbers are determined based on the findings from the previous section. We observe that beyond a certain threshold of real volumes, the improvement in DSC reaches a plateau.  This observation can be attributed to the limited number of modes covered by the synthetic images, which have been trained on only 30 real volumes, resulting in a narrower coverage of possible variations. Consequently, further augmenting the dataset with additional synthetic images does not yield significant benefits to the model's performance; a similar trend is observed with traditional data augmentation techniques. These results underscore the potential benefits of combining traditional data augmentation methods with generative model-based augmentation. Table \ref{tab:volumes} also highlights the challenges associated with training PET segmentation using a very limited dataset (less than 30 volumes), resulting in mediocre performance. This indicates that the model encounters difficulties in generalizing effectively to this modality. In contrast, the BRATS dataset demonstrates higher scores for a same data quantity. This further supports the earlier assertion regarding the challenges of discerning tumors in PET scans compared to FLAIR MRIs, where tumors are of significant size and well-defined characteristics.

\begin{table}[t]
\footnotesize
\centering
\resizebox{\columnwidth}{!}{%
\begin{tabular}{@{}clccccc@{}}
\toprule
\multicolumn{2}{c}{\multirow{2}{*}{Methods}} & \multicolumn{5}{c}{Number of training volumes} \\ \cmidrule(l){3-7} 
                         &                         & 10 & 20 & 30 & 40 & 50 \\ \cmidrule(r){1-2} \cmidrule(lr){3-3} \cmidrule(lr){4-4} \cmidrule(lr){5-5} \cmidrule(lr){6-6} \cmidrule(l){7-7}
\multirow{3}{*}{BRATS}   & 2D Reference               & 0.219$\pm$0.04 & 0.471$\pm$0.02 & 0.581$\pm$0.04 & 0.652$\pm$0.01 & 0.684$\pm$0.01 \\
                         & Ours ($\alpha = 0.01$)  & 0.666$\pm$0.02 & 0.687$\pm$0.01 & 0.724$\pm$0.01 & 0.724$\pm$0.00 & 0.721$\pm$0.00 \\ 
                         & Ours + Aug. ($\times$5) & \textbf{0.728$\pm$0.01} & \textbf{0.743$\pm$0.00} & \textbf{0.750$\pm$0.00} & \textbf{0.764$\pm$0.00} & \textbf{0.765$\pm$0.00} \\ \cmidrule(r){1-2} \cmidrule(lr){3-3} \cmidrule(lr){4-4} \cmidrule(lr){5-5} \cmidrule(lr){6-6} \cmidrule(l){7-7}
\multirow{3}{*}{HECKTOR} & 2D Reference               & 0.051$\pm$0.12  & 0.233$\pm$0.05  & 0.428$\pm$0.03  & 0.500$\pm$0.02  & 0.542$\pm$0.01  \\
                         & Ours ($\alpha = 0.01$)  & 0.460$\pm$0.01  & 0.540$\pm$0.01  & 0.594$\pm$0.00  & 0.584$\pm$0.01  & 0.588$\pm$0.00  \\
                         & Ours + Aug. ($\times$5) & \textbf{0.558$\pm$0.02}  & \textbf{0.609$\pm$0.01}  & \textbf{0.631$\pm$0.00} & \textbf{0.644$\pm$0.00} & \textbf{0.647$\pm$0.00}  \\ \bottomrule
\end{tabular}%
}
\caption{Quantitative performance of the our approach evaluated in terms of DSC (\%↑)  with respect to the base number of real volumes used for training. We also present an augmented variant of our approach in conjunction with standard augmentation methods, resulting in additional performance enhancement. \label{tab:volumes}}
\end{table}

Finally, we present the compiled segmentation results for a set of state-of-the-art architectures in the following Table \ref{tab:sota_segs}. Initially, their performance was evaluated using the limited dataset of only 30 volumes, which was then augmented using our method. Table \ref{tab:sota_segs} showcases the learning capabilities of specific architectures in data-scarce regimes. Notably, SwinUNETR achieved the highest reference score, attributed to its shifted windowing mechanism which enables improved attention scaling for small datasets. After the augmentation, the results tend to stabilize across most architectures. UNETR demonstrates the lowest results, as it is based on Vision Transformers \cite{dosovitskiy2020image} architecture known for its requirement of large datasets for convergence. Among these methods, nnUNet, with its default optimization settings, yields the best results, reaching a peak of 74.5\% DSC on BRATS and 61.0\% on HECKTOR.

\begin{table}[t]
\centering
\resizebox{\columnwidth}{!}{
\begin{tabular}{@{}lcccc@{}}
\toprule
\multirow{2}{*}{Methods}        & \multicolumn{2}{c}{BRATS} & \multicolumn{2}{c}{HECKTOR} \\ 
                                & Reference 3D & Our Aug. & Reference 3D & Our Aug.\\
\cmidrule(r){1-1} \cmidrule(lr){2-3} \cmidrule(l){4-5}
U-Net \cite{ronneberger2015u} & 0.607$\pm$0.01 & 0.724$\pm$0.01 & 0.454$\pm$0.03 & 0.594$\pm$0.00\\
UNETR \cite{hatamizadeh2022unetr}     & 0.506$\pm$0.01 & 0.708$\pm$0.01 & 0.308$\pm$0.03 & 0.478$\pm$0.01 \\
SwinUNETR \cite{hatamizadeh2021swin}  & \textbf{0.676$\pm$0.00} & 0.744$\pm$0.00 & \textbf{0.517$\pm$0.01} & 0.587$\pm$0.00 \\
Attn U-Net \cite{oktay2018attention} & 0.569$\pm$0.01 & 0.723$\pm$0.00 & 0.290$\pm$0.03 & 0.548$\pm$0.01 \\
nnU-Net \cite{isensee2018nnu}  & 0.618$\pm$0.00 & \textbf{0.745$\pm$0.00} & 0.452$\pm$0.01 & \textbf{0.610$\pm$0.00} \\
\bottomrule
\end{tabular}%
}
\caption{Quantitative performance comparison of state-of-the-art segmentation architectures in a data-scarce regime and after augmentation using our method. Evaluation based on DSC (\%↑). \label{tab:sota_segs}}
\end{table}

% Although the improvement gained with more real volumes may be relatively modest, the use of these approaches becomes a significant asset when dealing with limited data from real-world scenarios, which may not be as well-preprocessed and annotated as the curated BRATS and HECKTOR datasets.

\section{Conclusion \label{sec5}}
In this study, we introduced a novel hybrid architecture incorporating a discriminative regularization with the HVAE model for data augmentation in medical image segmentation tasks. The experimental results show that our proposed method outperforms traditional data augmentation techniques and state-of-the-art generative models in terms of DSC. Our future research involve expanding our suggested method to a new version of VAEs inspired from quantum mechanics. Built upon the Hamiltonian principle, we plan to explore the idea of viewing the latent space as a complete density rather than a single point in space, and further investigate how this approach can improve the generation of diverse, high-quality medical images.
%%%%%%%%%%%%%%%%%%%%%%%%%%%%%%%%%%%%%%%%%%%%%%%%%%%%%%%%%%%%%%%%%%%%%%%%%%%%%%%%%%%%%%%%%%%%

% Please add the following required packages to your document preamble:
% \usepackage{booktabs}
% \usepackage{multirow}
% \usepackage{graphicx}

% Please add the following required packages to your document preamble:
% \usepackage{multirow}

%% The Appendices part is started with the command \appendix;
%% appendix sections are then done as normal sections
%% \appendix

%% \section{}
%% \label{}

%% If you have bibdatabase file and want bibtex to generate the
%% bibitems, please use
%%
\bibstyle{elsarticle-harv} 
\bibliography{main}

%% else use the following coding to input the bibitems directly in the
%% TeX file.

% \begin{thebibliography}{00}

% %% \bibitem[Author(year)]{label}
% %% Text of bibliographic item

% \bibitem[ ()]{}

% \end{thebibliography}
\end{document}